\documentclass[fleqn,10pt]{wlscirep}
\title{Delineating elastic properties of kinesin linker and their sensitivity to point mutations}

\author[1,*]{Micha{\l} {\'S}wi\c{a}tek}
\author[2]{Ewa Gudowska-Nowak}
\affil[1]{Jagiellonian University, Marian Smoluchowski Institute of Physics, ul. Prof. S.Łojasiewicza 11,  Krak\'ow, 30--348, Poland}
\affil[2]{Jagiellonian University, Marian Smoluchowski Institute of Physics and Mark Kac Center for Complex Systems Research, ul. Prof. S.Łojasiewicza 11, Krak\'ow, 30--348, Poland}
\affil[*]{mgj.swiatek@uj.edu.pl}

\begin{abstract}
{We analyze free} energy estimators from simulation trials mimicking single-molecule pulling experiments on a neck linker of a kinesin motor. For that purpose, we have performed a version of steered molecular dynamics (SMD) calculations. The sample trajectories have been analyzed to derive distribution of work done on the system. In order to induce unfolding of the linker, we have stretched the molecule at a constant pulling force and allowed for a subsequent relaxation of its structure. The use of fluctuation relations (FR) relevant to non-equilibrium systems subject to thermal fluctuations allows us to assess the difference in free energy between stretched and relaxed conformations. 
To further understand effects of potential mutations on elastic properties of the linker, we have performed similar {\it in silico} studies on a structure formed of a polyalanine  sequence (Ala-only) and on three other structures, created by substituting  selected types of amino acid residues in the linker's sequence with alanine (Ala) ones.
The results of SMD simulations indicate a crucial role played by the Asparagine (Asn) and {Lysine (Lys)} residues in controlling stretching and relaxation properties of the linker domain of the motor.
\end{abstract}
\begin{document}

\flushbottom
\maketitle
%
%
\thispagestyle{empty}

\section*{Introduction}

The motor proteins' ability to generate movement creates a situation, where it is possible to treat different parts of one molecule as roughly independent objects, that possess an ability to move at different times and speeds \cite{yildiz2008intramolecular}. That makes the field of motor proteins a desirable testing ground for application of various theoretical models, that aim to filter the inherent complexity of biological systems \cite{kolomeisky2013motor,kolomeisky2005dynamic,teimouri2015theoretical,hyeon2011structural}. Among the motors, members of the kinesin protein superfamily are consistently used throughout the years in research focusing on molecular motors' mechanical properties, both {\it in vitro} \cite{kolomeisky2013motor} and {\it in silico} \cite{kolomeisky2013motor,teimouri2015theoretical,zhang2012dissecting,hyeon2011structural}.


Strain through the neck linker ensures processive runs of the motor \cite{Howard} and can be estimated by analyzing elastic properties of border regions between heads of the kinesin molecule. Here, in a desire to simplify a polymer-like model of the linker, we have neglected long range interactions and considered solely the structure of the {\it neck linker} itself. 
The {\it neck linker} label refers to a concise (less than 20 amino acids length) amino acid sequence in a single kinesin head that acts as a bridge between $\alpha$-6 helix in coiled-coil dimerization domain and $\alpha$-7 helix in the core motor domain, respectively \cite{hariharan2009insights}. An on-going accumulation of experimental data evidence suggests that a transition of the neck linker from a disordered (random coil) state to an ordered ($\beta$-sheet) conformation is a key factor in determining a mechanism of force-generation that is a crucial element of molecular motors' ability to move along microtubules \cite{hariharan2009insights}.

Biopolymers forming chains are often interpreted in terms of numerous approximations, all having roots in theoretical assumptions that form a basis of the Freely Jointed Chain (FJC) model. Among these, the worm-like-chain (WLC) model \cite{Bouchiat} seems to be the most relevant when it comes to describing a bending process of a semi-rigid biostructure, even though it has received some criticism \cite{Kutys,padinhateeri2013stretching}. Stretching of a peptide requires an application of a certain force and the relation between that force $F$ and the stable extension $x$ of the chain can be formulated as 

\begin{equation}
F=\frac{k_BT}{L}\left[\frac{1}{4}\left (1-\frac{x}{L_c}\right )^{-2}+\frac{x}{L}-\frac{1}{4}\right ]
\end{equation}

where $k_B$ is Boltzmann constant, $T$ stands for temperature, $L_c$ is the contour length of the polymer and $L$ its persistence length.
In a previous work, we have already presented a preliminary venture into the matter at hand, showcasing a difference in stretching process between a {\it neck linker} and an Ala-only polypeptide \cite{lisowski2012understanding}. Here, employing the methods of Molecular Dynamics and Normal Mode Analysis, we intend to deliver a more comprehensive description of a  possible relation between {\it neck linker's} amino acid sequence and the specificity of its function.
The paper is organised as follows: After a brief {\it Introduction}, the section {\it Material and Methods} presents basic methodology of the domain analysis and describes the setup of Molecular Dynamics (MD) simulations.
A number of theoretical considerations are discussed, pertaining to thermodynamic description of an amino acid chain, ability to determine its elasticity via the force-extension relations and significance of non-equilibrium dynamics as used in our simulations. All these are placed in subsections of their own. Next, in {\it Results and Analysis}, data collected in series of simulations are presented and examined. The last section, {\it Conclusions}, contains our closing remarks in which we summarise findings and highlight points of interest for future research in this field.

\section*{Methods}

\vspace{5 mm}
\noindent
\textbf{Domain analysis of kinesin heads}
\vspace{5 mm}

In the initial part of our studies we have identified dynamic domains in the structure of {\it Kinesin Heavy Chain} (taken from PDB Databank( id:3kin)\cite{kozielski1997crystal}) and analyzed deformations (low-frequency domain motions) which have been obtained with a simplified mechanical model proposed by Hinsen \cite{Hinsen}. The method is essentially based on observation that the low frequency modes describing motion of domains in proteins are influenced by anharmonic effects and in realistic environments become strongly overdamped, thus independent of the applied force field details.

The structure has been analyzed with {\it DomainFinder} application \cite{Hinsen1999}. Firstly, an approximate normal mode analysis has been performed. Then, 16 modes of lowest frequencies have been stored for further domain and deformation analysis. Energy threshold of 200 kJ/mol has been chosen to discriminate regions of sufficient rigidity to be candidates for domains. This choice has led to acquiring  images presented in FIG.\ref{domeny}, in which domains are associated with internally stable regions of protein and off-domain regions are relatively fluid.

Here, blue color represents regions of deformation energy well below threshold, while light blue and light red parts symbolize regions slightly below and slightly above the threshold, respectively. A crucial parameter differentiating between rigid regions with uniform motions and intermediate regions whose internal deformation yields systematic contributions to the overall motion between boundaries of the domain is the
domain coarseness factor $c$. In brief, the coarseness parameter specifies how similar the rigid body motions of different residues should be to consider that these residues form a dynamical domain; thus the smaller the coarseness, the finer is the definition of the domains. Since the deformation energies of chosen 16 modes have ranged up to 15-20,  the deformation analysis based on these 16 modes has been performed with a deformation threshold equal to 15. The dynamical domain decomposition has resulted in images displayed in the lower panel of FIG.\ref{domeny}.

A deformation energy definition used by {\it DomainFinder}  relates to the interaction energy between two particles $i$ and $j$ of the elastic network and reads
\begin{equation}
E_i=\frac{1}{2}\sum^N_{j=1}k(R_{ij}^{(0)})\frac{|(d_i-d_j)R_{ij}^{(0)}|^2}{|R_{ij}^{(0)}|^2}
\end{equation}
where $i, j$ denote $C_{\alpha}$ atoms, $d_{i,j}$ are corresponding infinitesimal displacements from original positions (displacement of the atom in the mode to be analyzed), $R_{ij}^0$  are distances between pairs of atoms $i, j$ in a submitted structure and $k(R_{ij}^0)$ is an effective harmonic force constant, that attenuates with a spatial distance according to the relation:
\begin{equation}
k(R_{ij}^{(0)})=C\exp\left ( -\frac{|R_{ij}^{(0)}|^2}{r_0^2}\right)
\end{equation}
in order to maintain a force field of short range ($r_0$), that excludes interactions between potential domains. Parameter $C$ has been chosen arbitrarily, as 47,400 kJ$\times$ mol$^{-1}$ nm$^{-2}$ at temperature 300K, to ensure
compatibility with the {\it Amber 94} forcefield. Amplitudes of $d_{i,j}$ are defined with an equation 
\begin{equation}
\sum^N_{i=1}|d_i|^2=fN,
\end{equation}
$f $ being a scaling factor of value 1 nm$^2$.
Displacements $d_i$ have been used to define rotation ($\Omega$) and translation ($T$) vectors:
\begin{equation}
d_i=T+\Omega\times R_i
\end{equation}
where $R_i$ stands for the $i$  atom position. When displacement vectors do not describe pure rigid-body motion, linear least-squares fit is used to determine values of $T$  and $\Omega$. 
The structure is further divided into cubic compartments of side length 1,2 nm, all ignored unless containing at least 3 atoms and having average deformation energy below pre-defined threshold. Those cubes have their rotation and translation vectors calculated. A following definition of similarity is used to identify clusters of cubes having similar mobility.
\begin{equation}
{\cal{S}}_{ij}=3\frac{|\Omega_i+\Omega_j|}{|\Omega_i-\Omega_j|}+\frac{|T_i+T_j|}{|T_i-T_j|}
\end{equation}
The rotation vectors are empirically more precise in sorting out domains, thus are given greater weight. A cluster is finally created, by using the criterion
\begin{equation}
{\cal{S}}_{ik}>\frac{{\cal{S}}_{ij}^{max}}{c}
\end{equation}
where $c$ is pre-selected domain coarseness parameter. All cubes contributing to this relation, are then considered to compose one cluster.

\vspace{5 mm}
\noindent
\textbf{Structure and relaxation of extended linker: mechanical and thermodynamic considerations}
\vspace{5 mm}

By definition, the partition function for simulations run at a constant volume condition is given by
\begin{equation}
Z={\cal{N}}\int d {\mathbf r}e^{-\beta E(\mathbf{r})}
\label{partitiona}
\end{equation}
where $\beta=k_BT$. Here  $T$ is absolute temperature and $E$ stands for the energy of a given configuration state, with an average energy of the system given by
\begin{equation}
\left < E\right >=\int d {\mathbf r} E(\mathbf{r})\rho(\mathbf{r}),
\end{equation}
in which $\rho(\mathbf{r})$ is the equilibrium probability density $\rho (\mathbf{r})=e^{-\beta E(\mathbf{r})}\times [\int  d {\mathbf r}e^{-\beta E(\mathbf{r})} ]^{-1}$.
Accordingly, the configurational entropy of the system is given by 
\begin{equation}
S=\frac{\left < E\right >-F}{T}=k_B\log {\cal{N}} - k_B\int d\mathbf{r} \rho(\mathbf{r})\log\rho(\mathbf{r})
\end{equation}
with $F$ being the Helmholtze free energy $F=-k_BT\ln Z$.
If the system energy is partitioned over many local minima (energy wells), the configurational integral Eq.\ref{partitiona} can be represented  \cite{Mei,Chong} in the form  of a sum 
$Z=\sum_i Z_i$ with $Z_i={\cal{N}}\int_i d {\mathbf r}e^{-\beta E(\mathbf{r})}$ and integral evaluated over the $i-th$ energy well, in which the probability density $\rho_i(\mathbf{r})$ can be expressed as
\begin{equation} 
\rho_i(\mathbf{r})=\frac{\rho(\mathbf{r})}{p_i}, \;\;\; \mathbf{r}\in \Omega_i
\end{equation}
with $p_i=\frac{Z_i}{Z}$ \cite{Hill}. The average  energy of the system can be then rephrased as $\left < E\right >=\sum_ip_i\left < E\right >_i$ with entropy 
\begin{equation}
S=-k_N\sum_i p_i\ln p_i+\sum_ip_i S_i
\end{equation}
given by the sum of weighted average of individual entropies $S_i$  associated with different wells and the entropy of partitioning of the system among various wells. All in all, the weighted average Eqs.(10) and (12), or the differences $\Delta S$, $\Delta F$ pertinent to two (initial/final) states can be then accessed in a straightforward way by a histogram method counting the number of times the molecule "visited"  given configurational state in course of MD simulations \cite{Mei}.

\vspace{5 mm}
\noindent
\textbf{MD simulations' setup}
\vspace{5 mm}

A structure of a motor domain, belonging to a kinesin-like protein {\it KIF3B} (a Kinesin-2 family's member) \cite{lawrence2004standardized}, has been obtained from the PDB Databank( id:3b6u)\cite{berman2000protein}. A sequence of 19 amino acids, 17 of which are considered to be a functional part known as a {\it neck linker}, has then been extracted and optimized. In order to do so, {\it Steepest Descent} and {\it Conjugate Gradients} algorithms have been employed.
The neck linker chain was subsequently placed in a box of water molecules (a Simple Point Charge model), and the whole system was optimized again. Next, a simulation of Molecular Dynamics (MD) was scheduled, with a goal of achieving a state of at least near equilibrium, producing a $6ns$ long trajectory (with a time step duration $\Delta t= 2\times 10^{-3}ps$). A distribution of {\it end-to-end} distances was then created and used to determine the mean {\it end-to-end} distance of the equilibrated linker chain. Finally, a structure has been selected with an {\it end-to-end} distance sufficiently close to the mean, while belonging to a time frame from near the end of the simulation.

That selected {\it neck linker} structure has been taken out of the box of water molecules and placed in the implicit solvent. A short simulation, with positions of first and last $C_{\alpha}$ atoms fixed, served as a short equilibration routine. After that, the system was employed as a starting point of 10$^4$ simulation runs, where a constant force of 1300 $kJ*mol^{-1}*nm^{-1}$ ( approximately 2160 $pN$) was applied between a mass centre of the 1st residue and a mass centre of the 19th residue. We desired to test our structure against strong external influence that causes a rapid response from our system and, with that in mind, the aforementioned force value has been chosen, after some preliminary MD runs. Those simulations of non-equilibrium dynamics produced a set of $1ps$ trajectories. The final states of these trajectories became starting points of another 10$^4$ simulations, each lasting $1ps$ ($\Delta t= 2\times 10^{-3}ps$), where the constant stretching force had been turned off, resulting in system relaxation. In all simulations, the {\it Berendsen} thermostat has been used to ensure stable temperature conditions (T=300K), while the levels of pressure have been controlled with {\it Parrinello-Rahman} barostat (p=100kPa).
Since our interest has been focused on investigation of linker's specific elasticity, analogous steps have been taken with regards to an 18 amino acid long {\it Ala-only} peptide, with the equilibration simulation being 3ns long. The alanine residue is the simplest possible, not possessing any side chain and, because of that, a polyalanine sequence has been deemed the best model for observing a protein polymer behavior limited only to the protein backbone.
Additionally, the above steps of modelling have been repeated for three "intermediate" structures between the original {\it neck linker} and the {\it Ala-only} polymer. Namely, by selecting some of amino acid residues (either Asparagine, or Proline, or Lysine)  and substituting them with alanine residues, "mutant" versions of the neck-linker have been created. The equilibrium simulations of those three altered linker structures lasted $5ns$, $6ns$ and $3ns$ for {\it no-Asparagine}, {\it no-Lysine} and {\it no-Proline} chains, respectively.
Finally, the whole process involving all 5 different sequences has been repeated for different stretching force values (130, 400, 700 and 1000 $kJ*mol^{-1}*nm^{-1}$).
The 4.5.5 version \cite{pronk2013gromacs} of the GROMACS package \cite{berendsen1995gromacs,lindahl2001gromacs} and the {\it OPLS-aa } force field \cite{jorgensen1988opls,jorgensen1996development} have been employed to perform all MD simulations.

\vspace{5 mm}
\noindent
\textbf{Steered molecular dynamics of kinesin linker structure}
\vspace{5 mm}

Molecular interactions and mechanical properties of individual molecules can be nowadays probed by use of combined techniques, like Atomic Force Microscopy (AFM) and optical tweezers \cite{Ritort}. In such experiments single molecules are hold and stretched, and from the measurements of a cantilever spring restoring force in the AFM instrument, the information about  elasticity (effective spring constants) and  intensity of rupture forces can be derived \cite{Szabo,Park}. Analogous to these experimental setups, steered molecular dynamics (SMD) simulations permit similar investigations to be performed {\it in silico}. In brief, the procedure of SMD applies external steering forces in molecular dynamics simulations to investigate  processes of e.g. protein unfolding or binding/unbinding of substrates separated by some energy barriers. Practical designs of such simulations are based on relating free energy difference in nonequilbrium steady states achieved in course of manipulation with the work done through the process.
The thermostated system is hold at the beginning of the action at equilibrium of a given temperature $T$. By changing an externally controlled parameter $\lambda$, the external work $W$ done on the system may be estimated. The process is repeated many times  so that the statistics of work performed is collected with free energy difference between the steady states related by Jarzynski equality \cite{Park,Jarzynski}
\begin{equation}
\left < e^{-\beta W}\right>=\int p(W)e^{-\beta W}=e^{-\beta \Delta G},\;\;\;\beta=(k_BT)^{-1}
\end{equation}
with average taken over repeated realizations of the process and $p(W)$ being the relevant probability density function (PDF) for work distribution. The microscopic state of the system composed of $N$ particles in contact with the heat bath is specified by $3N$-dim position vector $\textbf r$ and $3N$-dim momentum vector $\textbf p$ evolving under the dynamics governed by the Hamiltonian $H(\textbf r,\textbf p)$. Description of the system dynamics in terms of a collective variable $ x$  (reaction coordinate) aimed to capture essentials of undergoing thermodynamic process is possible by defining a potential of mean force $V(x)$ expressed as
\begin{equation}
e^{-\beta V(x)}=\int d{ \textbf r}d{\textbf p}\delta(x-x'({ \textbf r}))e^{-\beta H(\textbf r,\textbf p)}
\end{equation}
If the system is further perturbed by an external potential $V_{\lambda}$ with a control parameter $\lambda$, the total potential energy of the system becomes $U(x,\lambda)=V(x)+V_{\lambda}(x)$ and the total Hamiltonian changes to $\tilde{H}_{\lambda}(\textbf r,\textbf p)+H'_{\lambda}$.  In the experiments with pulling force, the center of mass of the pulled molecule is attached to a spring with an elasticity constant $k$, so that the pulling force is $F=k(\lambda(t)-x)$ with the control parameter $\lambda(t)=x_0+vt$. The thermodynamic work definition for overdamped dynamics becomes then 
\begin{equation}
W=\int dt \dot{\lambda}\frac{\partial U(x;\lambda)}{\partial \lambda}\equiv-\frac{k}{\gamma}\int^t_0du(x_u-vu)
\end{equation}
where $\gamma$ stands for the friction coefficient.

The implications of non-equilibrium dynamics described above have been taken into consideration as we set out to gauge the free energy difference between stretched and relaxed amino acid chains.

\section*{Results and Discussion}

\begin{figure}[!htb]
\begin{tabular}{c}
\includegraphics[angle=0,width=0.9\columnwidth]{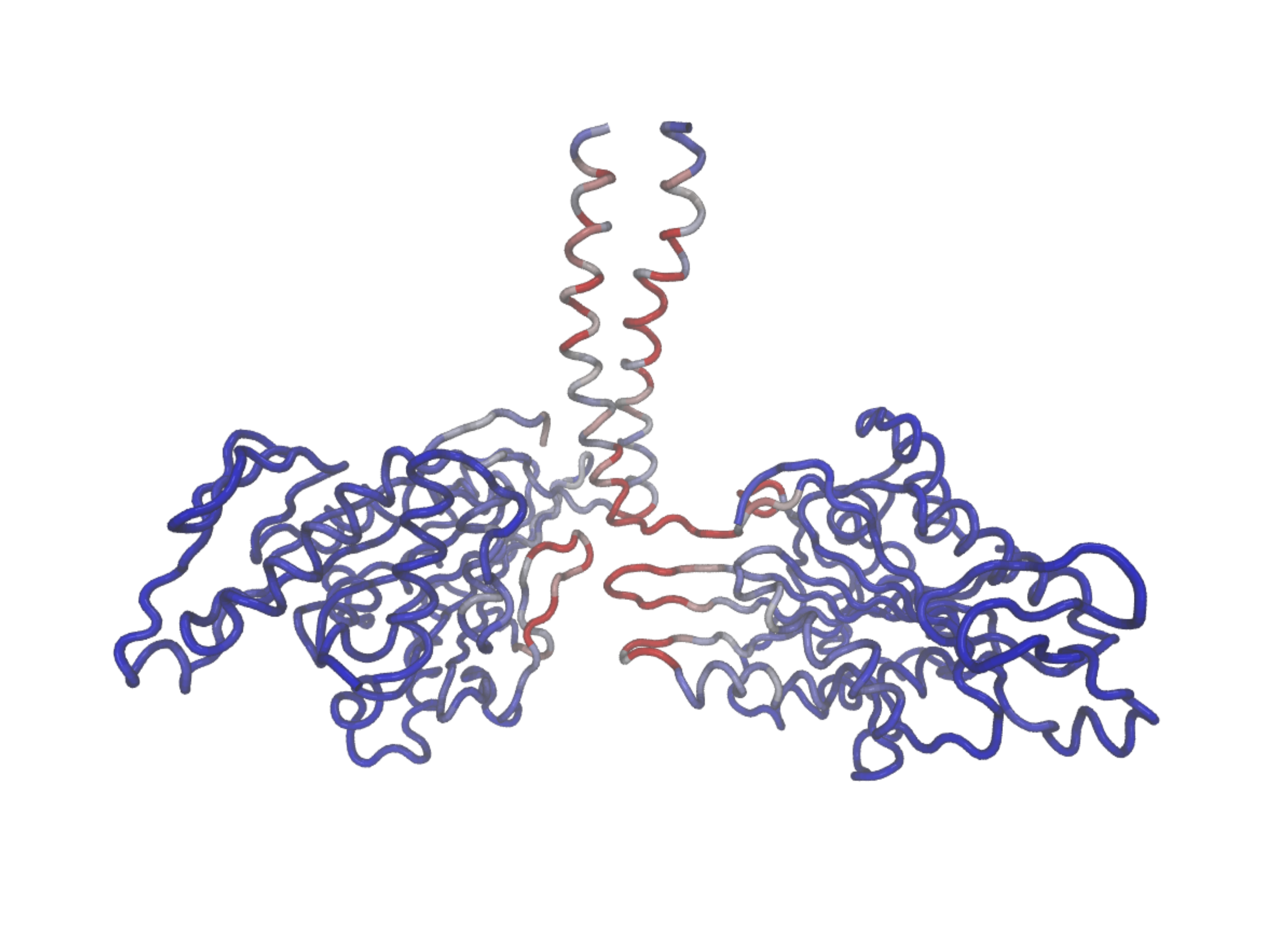}\\
\includegraphics[angle=0,width=0.9\columnwidth]{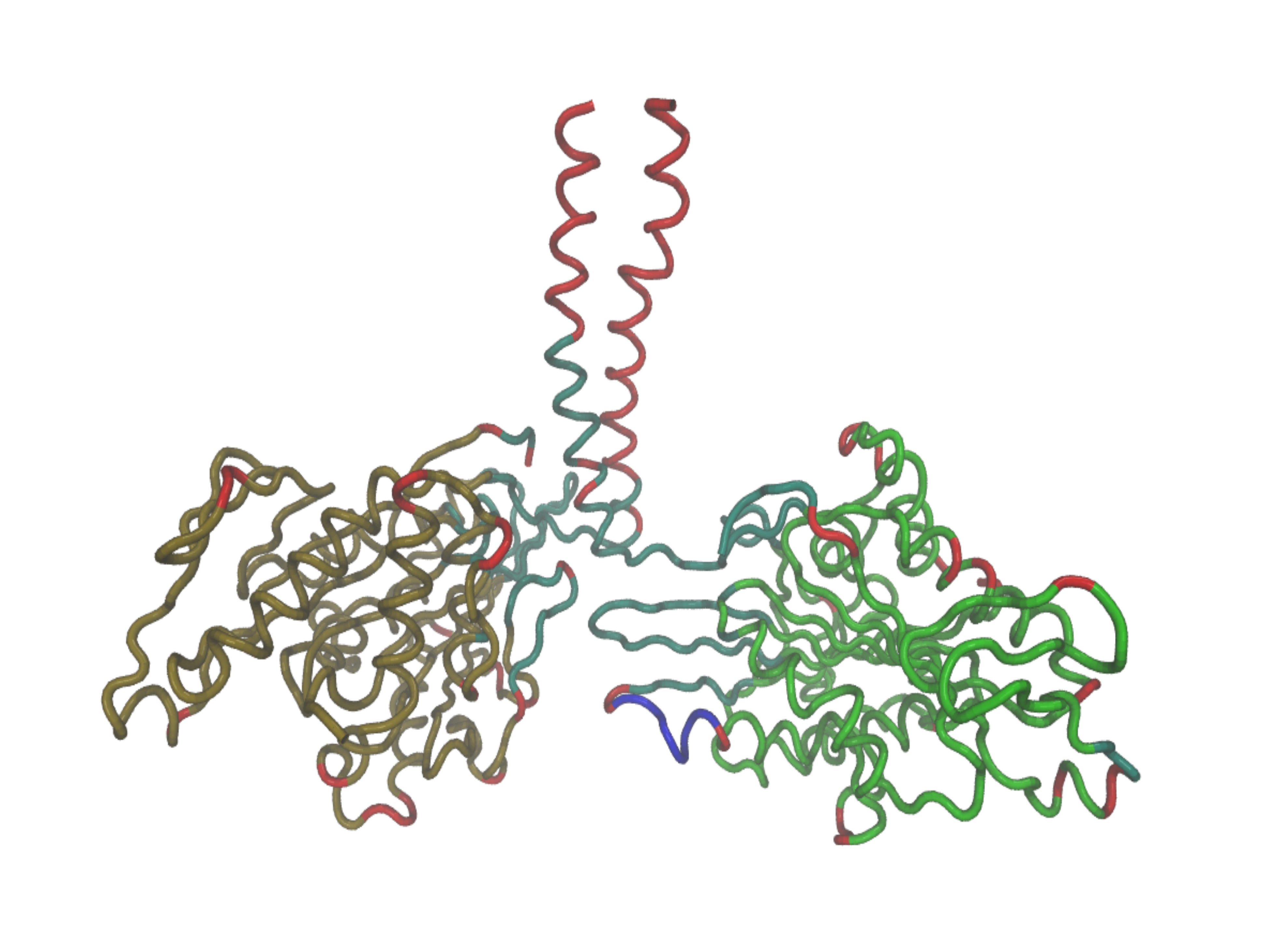}
\end{tabular}
\caption{
Distribution of deformation energies in the kinesin structure (upper panel) and derived domains (lower panel). Clusters of vibrational energies have been plotted by similar colors.}
\label{domeny}
\end{figure}

{A clear decomposition of the kinesin structure to three domains comprising two heads can be seen in FIG.\ref{domeny},}with softest motions observed in the region of the neck linker. This domain formed of 14-18  amino acids is widely considered a key structure  underlying kinesin's force-generating mechanism and has been examined in a series of experimental \cite{Block} and theoretical \cite{hariharan2009insights} studies. In particular, it has been proposed that a conformationally flexible unstructured state of the linker changes to a structured and docked one upon ATP binding, providing essential conformational change in the motor, responsible for subsequent stepping \cite{Block}. On one hand, the linker spring has to be then flexible enough to allow for diffusive search of the motor head of the next binding side. On the other, when both heads of the motor are simultaneously bound to the microtubule track, the neck linker has to be sufficiently stiff to ensure that mechanical forces between both head domains enable mechanical coupling. Accordingly, mechanical models and molecular dynamics simulations of this peptide structure are important contributions to understanding its elastic properties and  ability to control kinesin's motion.

In order to select  a proper starting structure for stretching simulations, \textcolor{black}{a construction of well equilibrated initial ensemble is required.}
It is often assumed that, for a small system with properly functioning temperature and pressure coupling, a run that does not exceed $100ps$ is enough to achieve a state of equilibrium \cite{hariharan2009insights,zhang2003molecular}. However, it has been shown that such assumption does not have to be necessarily true \cite{genheden2012will,smith2002assessing}. Taking into consideration possible difficulties in achieving equilibrium, we have decided to measure fluctuations of the {\it end-to-end} distance, aside from a routine check of parameters (e.g. Root Mean Square Displacement (RMSD) or average potential energy), typically used as indicative  measures  for an equilibrated system.

\begin{figure*}[htbp]
\begin{center}
\centering
\resizebox{1.1\columnwidth}{!}{
\includegraphics[width=\linewidth]{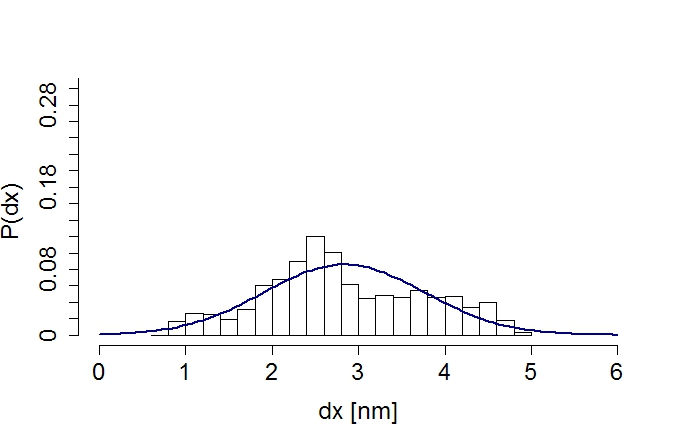}
\includegraphics[width=\linewidth]{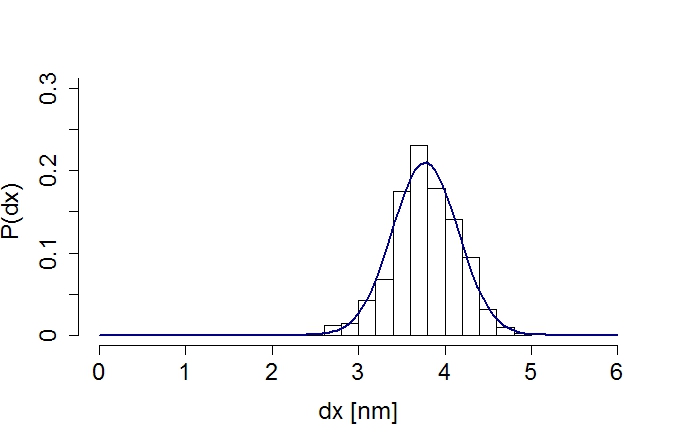}
} 
\resizebox{1.1\columnwidth}{!}{
\includegraphics{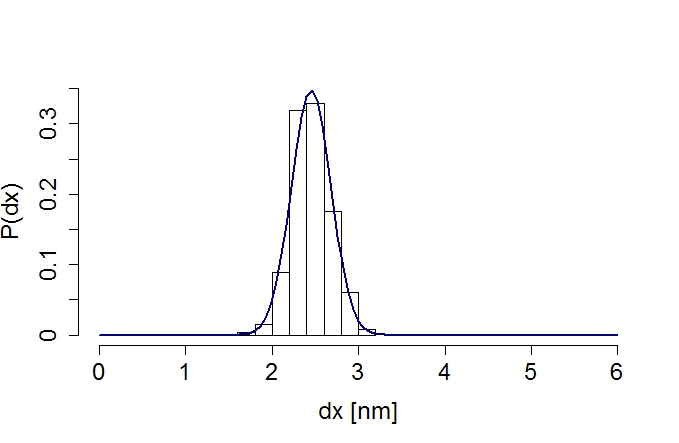}
\includegraphics{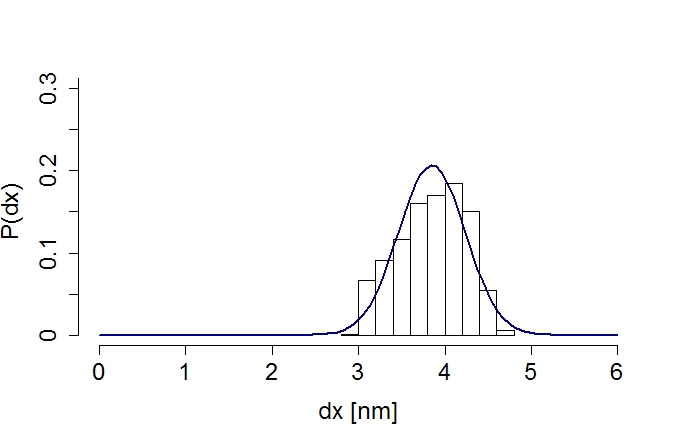}
}
\resizebox{1.1\columnwidth}{!}{
\includegraphics{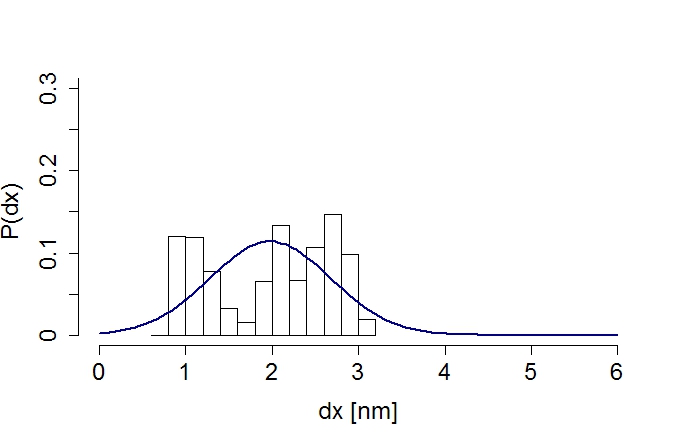}
\includegraphics{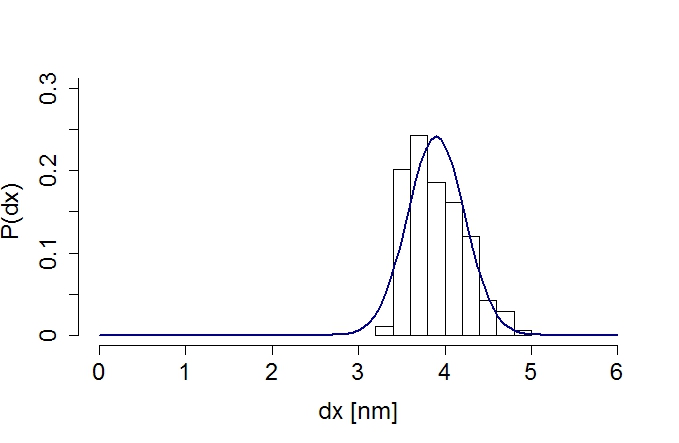}
}
\caption{Left column includes (from top to bottom): the {\it neck linker's} {\it end-to-end} distance distributions of the whole equilibrium simulation, 10$^4$ recorded steps preceding the last 10$^4$ and, in the final row, the last 10$^4$ recorded steps. Right column includes the {\it Ala-only} peptide's {\it end-to-end} distance distributions arranged in an analogous way. Gaussian curves have been fitted to guide an eye, based on the mean and $\sigma^2$ of a given empirical distribution derived from the SMD simulations.}
\label{CLT}
\end{center}
\end{figure*}

Figure \ref{CLT} displays {\it end-to-end} distance distributions in different time windows of the simulation with Gaussian curves  fitted to the data. \textcolor{black}{For a chain made up  of orientationally uncorrelated (free-jointed) links with a length of each segment randomly distributed, the {\it end-to-end} stretch distance is expected to follow statistics of  the Gaussian law. Although we have observed that the length distributions of the {\it neck linker} as well as the {\it Ala-only} polypeptide do fit  Gaussian curves in certain time regimes, the long-run equilibrium simulations of the segments clearly indicate deviations of the {\it end-to-end} distances from Gaussianity, see FIG.\ref{CLT}. This observation stays in line with the assumed GROMACS force field, which apart from the harmonic approximations on bonds and angles, contains also  
long range interactions:}

\begin{eqnarray}
V=\sum_{bonds}\frac{k_i}{2}\left(x_i-x_{i,0}\right)^2+\sum_{angles}\frac{k_i}{2}\left(\theta_i-\theta_{i,0}\right)^2 \nonumber + \sum_{torsions}\frac{V_n}{2}\left(1+cos\left(n\omega-\gamma\right)\right) \nonumber \\ + 
\sum_{i}^N\sum_{j=i+1}^N\left(4\epsilon_{ij}\left[\left(\frac{\sigma_{ij}}{r_{ij}}\right)^{12}-\left(\frac{\sigma_{ij}}{r_{ij}}\right)^6\right]+\frac{q_iq_j}{4\pi\epsilon_0r_{ij}}\right).\nonumber\\
\end{eqnarray}

Here $x_i$ is a symbol of an $i^{th}$ bond length, $\theta_{i}$ stands for an $i^{th}$ angle, while $x_{i,0}$ and $\theta_{i,0}$ are their respective reference values. $V_n$ is a parameter that gives information about rotation barriers of a torsion angle $\omega$, while $k_i$ refers to an $i^{th}$ force constant. $\epsilon_{ij}$ is a minimal value of Van der Waals potential between atoms with indices $i$ and $j$, $r_{ij}$ is a distance between these atoms, $\sigma{ij}$ is a distance between them when the Van der Waals potential value equals 0. The symbols $q_i$ and $q_j$ refer to charges of $i^{th}$ and $j^{th}$ atom respectively, and $\epsilon_0$ stands for the dielectric constant.

Simulation results indicate that when the chain's structure attains a local minimum of the potential, the long range interactions do not play a significant role. Effectively, their influence on variations of the potential energy wanes temporarily. As a result, the chain is able to explore a narrow conformational subspace, behaving similarly to the Gaussian chain, before being pulled out of the energy well by thermal fluctuations. If the chain never leaves the vicinity of that particular energy well, an overall distribution of {\it end-to-end} distances approaches a normal distribution for sufficiently long simulation runs.

\begin{figure*}[htp]
\begin{center}
\resizebox{1\columnwidth}{!}{
\includegraphics{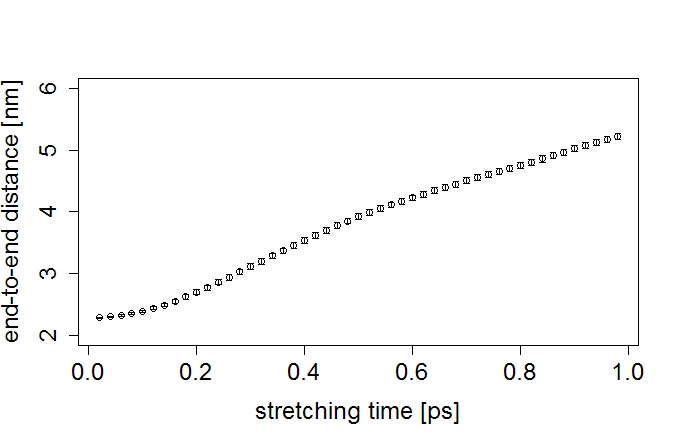}
\includegraphics{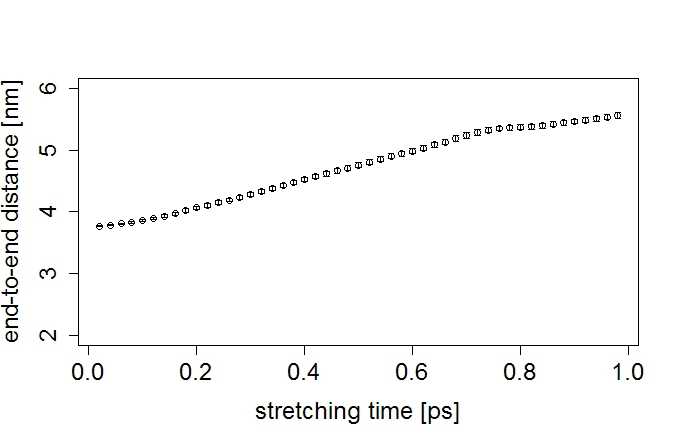}
} 

\caption{
{\it End-to-end} distances as functions of time, averaged over 10$^4$ MD runs, each of duration $t=1ps$. The left plot depicts a change in the {\it neck linker's} {\it end-to-end} distance. The right plot displays analogous findings for the {\it Ala-only} chain.
}
\label{t_scale}
\end{center}
\end{figure*}

{The mean {\it end-to-end} distance of the {\it Ala-only} chain over the whole simulation equals 3.77 $\pm$ 0.38 nm, hence such an equilibrated sequence has been chosen to be a starting point of the stretching process. In contrast, distribution of {\it end-to-end} distances of the {\it neck linker} structure is not as well fitted to a Gaussian curve. In order to make sure that a chosen structure is sufficiently close to the potential energy minimum, we have selected a model one  with the {\it end-to-end} distance of $2.27nm$, belonging to the class of conformations attained between $4.4$ and $5.2ns$ of simulation runs.  Results of the pulling experiments performed on chosen {\it neck linker} and {\it Ala-only} chains are displayed in FIG.\ref{t_scale} and clearly indicate that both structures stretch at (almost) a constant rate for the majority of the process. 
At the same time, in accordance with findings reported in our prior studies \cite{lisowski2012understanding}, we observe that {\it Ala-only} chain's linear response drops much sooner than that of the {\it neck linker} structure.}

\begin{figure*}[htp]
\begin{center}
\resizebox{1\columnwidth}{!}{
\includegraphics{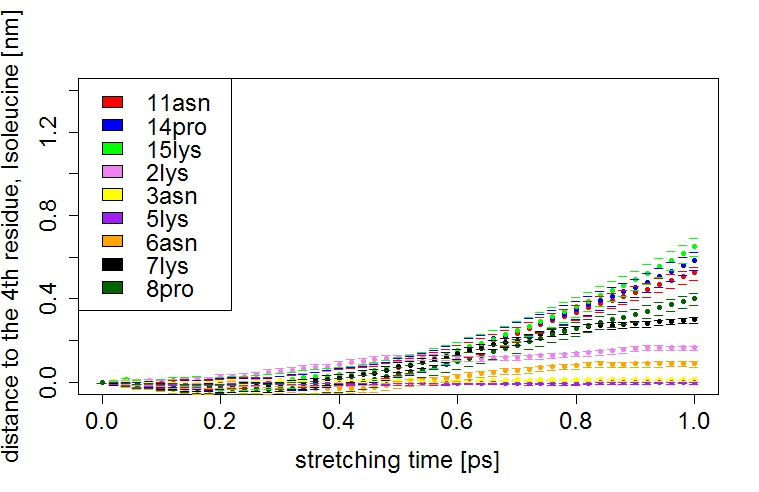}
\includegraphics{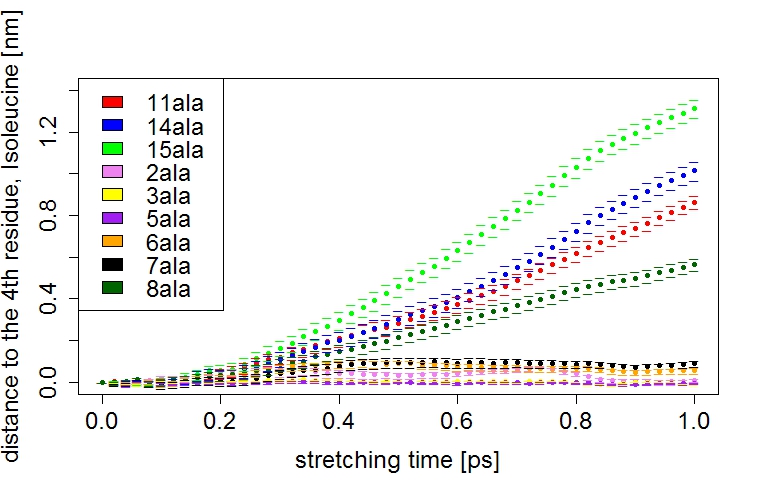}
} 
\caption{
Distances (extensions)  between residues as functions of time, averaged over 10$^4$ MD runs, each of duration $t=1ps$. The left plot depicts different rates of distance changes in the {\it neck linker's} chain between the 4th residue and selected residues, listed in the inset. The right plot displays
analogous findings  for the {\it Ala-only} chain.
}
\label{internal_d}
\end{center}
\end{figure*}

While the {\it end-to-end} distance is a useful parameter in AFM experiments and those mimicking them {\it in silico}, it does not give detailed information regarding inner dynamics of the examined structure. In order to gather additional information that could hint at inner dynamic characteristics and elasticity of the analyzed biopolymer chains, we have measured pair distances between the 4th residue of the simulated chains and a selected residue of interest. The 4th residue of our neck linker sequence is an Isoleucine amino acid which is one of the most prevalent elements at this position in a {\it neck linker} sequence across numerous kinesin families\cite{hariharan2009insights}. The second selected residue in the pair has been chosen as either neighboring  Asparagine, Lysine or Proline. Asparagine and Lysine have been chosen for their significant propensity to be in contact with water (polar Asparagine and positively charged Lysine), while Proline - because of the presence of a Pyrrolidine, five-member ring in its side chain, being the only steric group of that kind in the whole {\it neck linker} chain. Effects of the simulation runs are displayed in FIG.\ref{internal_d} and document considerable difference in response to mechanical perturbations between the {\it neck linker} and the {\it Ala-only} chain.

The extensions between the 4th residue and its proximal contacts (the 2nd, 3rd, 5th and 6th residues in the chain)  remain relatively unchanged throughout the stretching time, regardless of the type of the representative polymer chain. Stronger variations are observed for more distant pairs:  for the {\it neck linker} structure significant changes in extension profiles between pairs of residues emerge in time windows longer than $0.6ps$. At the same time the {\it Ala-only} sequence shows pronounced variability in conformations by comparison to a much more rigid structure of the {\it neck linker}.

Positions of the 8th and 14th residues in the {\it neck linker} are taken by Proline which is known to reduce flexibility of the chain in the kink (cis) conformation. In fact, in former {\it in silico} studies examining mechanical properties of the neck linker domain from sequence analysis \cite{hariharan2009insights} it has been argued that the cis-trans isomerization of a conserved proline residue switching between straight and kink forms accounts for variations in resulting force-extension profiles and supports experimental observations \cite{lu2007prolyl} of the proline's isomerization influence on duration and effectiveness of biological processes dependent on protein folding.                                                                                                                                                

In case of the {\it Ala-only} chain, displacement patterns between Isoleucine at 4th position and subsequent Alanine residues differ significantly from those observed for the {\it neck linker} chain stretched at constant pulling speed: In course of pulling experiment inner distances do not deviate much from their averages, whereas distances to external residues (at 8th, 11th, 14th and 15th positions) show pronounced extensibility. Altogether, while the final {\it end-to-end} distance of the {\it Ala-only} chain has been on average smaller than that of the {\it neck linker} (see FIG.\ref{t_scale}), its inner extension distances reach greater lengths, to the point, where 15th residue of the polyalanine chain has almost twice the final value, when compared to the largest inner distances of the {\it neck linker}. This indicates that stretching of the {\it Ala-only} chain is far more complex, possibly with emergent inner dynamic domains facilitating extensions.

\begin{figure}[!htb]
\centering
\includegraphics[angle=0,width=0.6\linewidth]{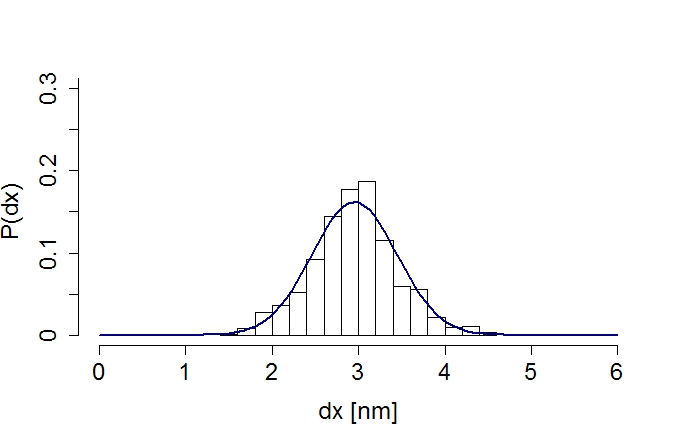} \\
\includegraphics[angle=0,width=0.6\columnwidth]{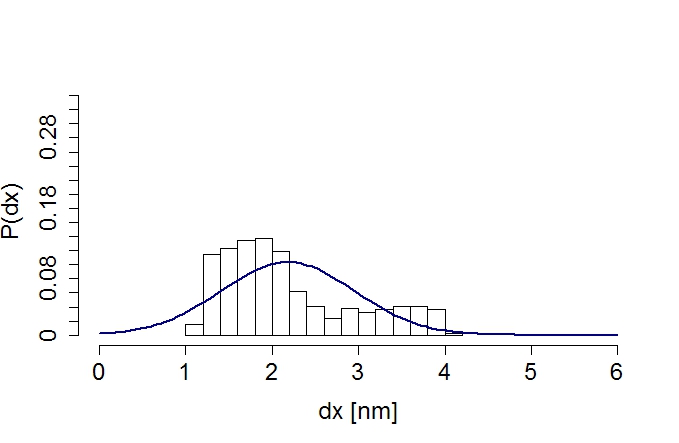} \\
\includegraphics[angle=0,width=0.6\columnwidth]{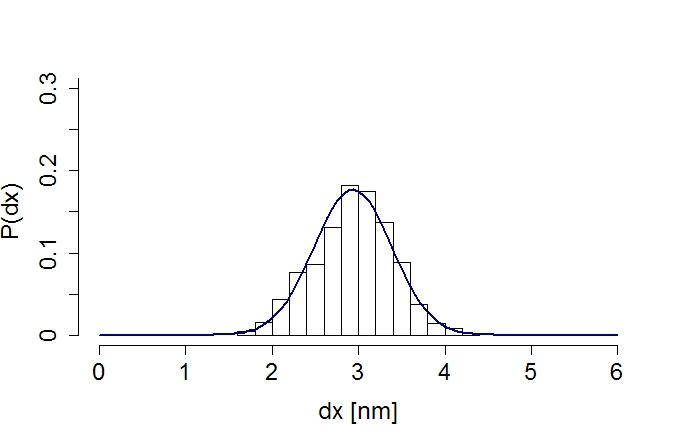} \\
\caption{
The {\it end-to-end} distance distributions of the {\it no-Asparagine} chain (top), the {\it no-Lysine} chain (middle) and the {\it no-Proline} chain (bottom), all pertaining to equilibrium simulations. Appropriate Gaussian curves have been fitted to the data, based on derived means and $\sigma^2$ of respective distributions.
}
\label{mutant_eq_ete_distr}
\end{figure}

\begin{figure}[!htb]
\centering
\includegraphics[angle=0,width=0.6\columnwidth]{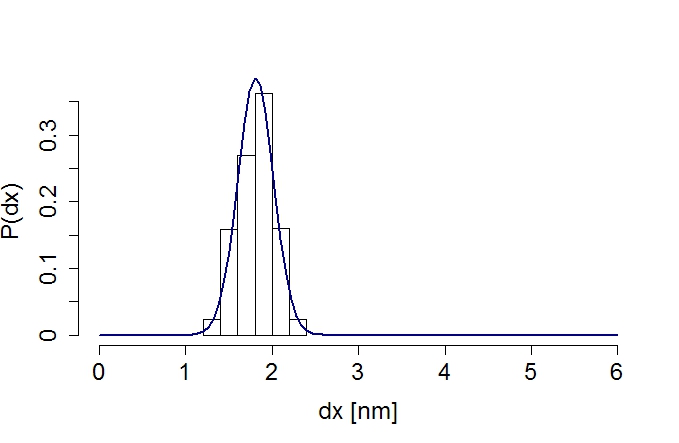} \\
\caption{
The {\it end-to-end} distance distributions of the last 10000 recorded steps taken from the {\it no-Lysine} peptide equilibrium simulation data. An appropriate Gaussian curve has been fitted to the data, based on the mean and $\sigma^2$ of its distribution.
}
\label{mutant_eq_ete_lys_distr}
\end{figure}

In order to further explore the specificity of {\it neck linker's} sequence and to determine how the presence of particular amino acid types affects that specificity, we have prepared 3 modified {\it neck linker} chains, where all Asparagine, all Lysine residues and the two Proline residues have been substituted with Alanine amino acids, respectively. Simulations' setup has been identical to the one employed in case of the unchanged {\it neck linker} and the {\it Ala-only} chain. The {\it no-Asparagine} and the {\it no-Proline} peptides seems to have easily achieved a local minimum of potential energy. On the other hand, the distribution of the {\it no-Lysine} chain {\it end-to-end} distances from the full simulation does not fit the Gaussian curve at all (FIG.\ref{mutant_eq_ete_distr}). The possible reasons for such behavior has been discussed above.
Accordingly, in order to meet the requirement of beginning simulation runs with a mechanically equilibrated structure, as a starting conformation of the {\it no-Lysine} chain we have selected a structure from a time period, in which the {\it no-Lysine} {\it end-to-end} distances have been distributed normally (FIG.\ref{mutant_eq_ete_lys_distr}).

\begin{figure*}[!htb]
\begin{tabular}{c}
\includegraphics[angle=0,width=1\columnwidth]{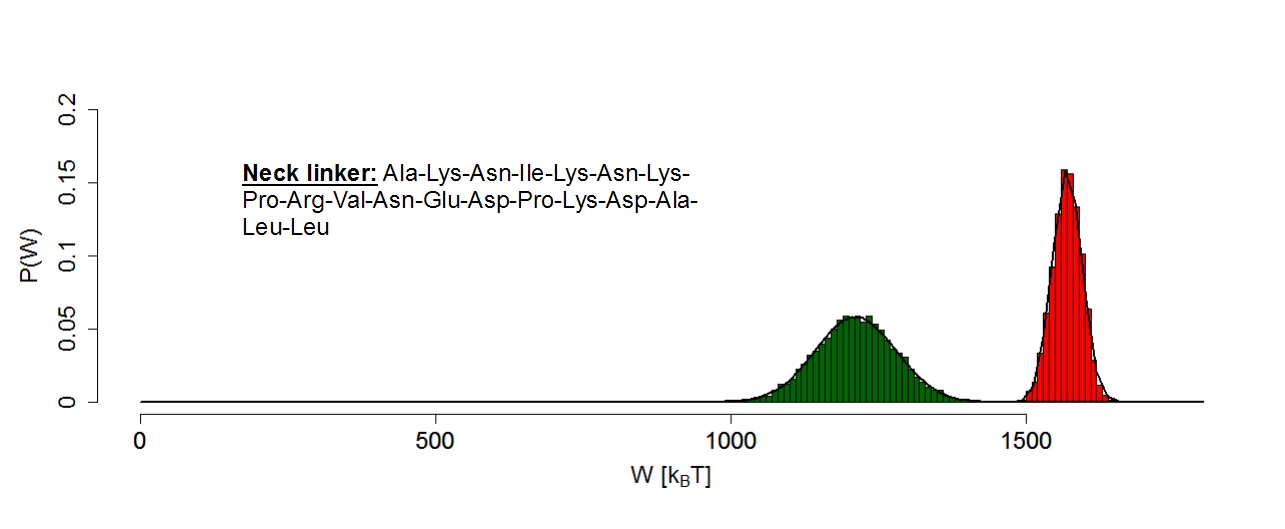}\\
\includegraphics[angle=0,width=1\columnwidth]{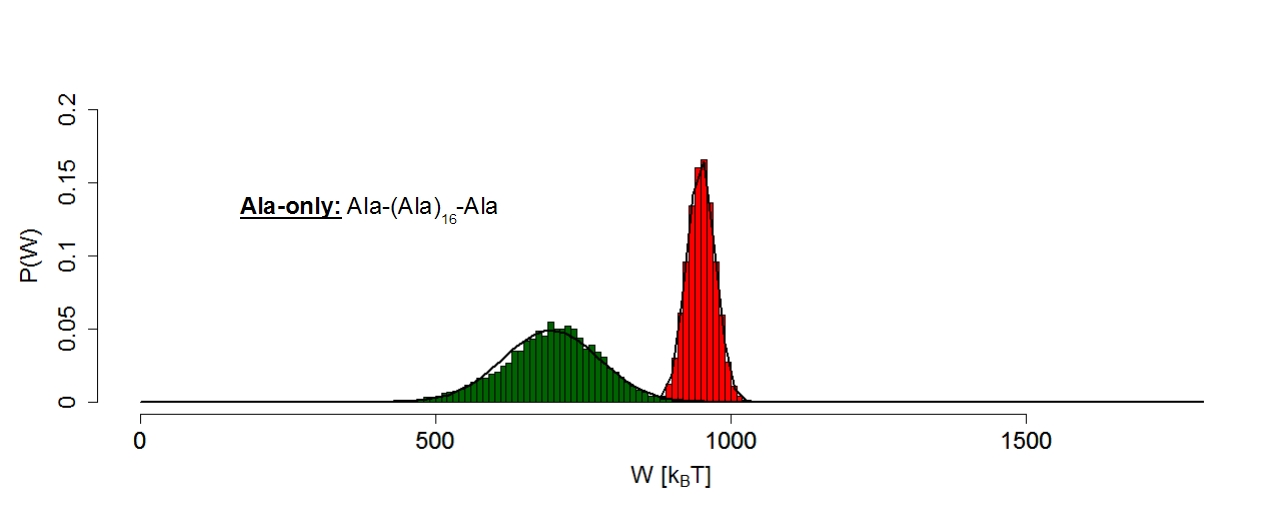}\\
\end{tabular}
\caption{
Reconstruction of probability density functions (PDFs) of work done on the {\it end-to-end} distance in pulling {\it in-silico} experiments of the modeled peptides (the stretching force value set at 1300 $kJ*mol^{-1}*nm^{-1}$). The Gaussian curves are fitted to the derived histograms of the {\it  neck linker} (top) and {\it Ala-only} peptide (bottom). Crossing points of the curves are at $1462.4 k_BT$ and $882 k_BT$, 
respectively. The ensembles of stretched structures are red, while the ensembles of relaxed structures are green colored.
}
\label{work_distr}
\end{figure*}

It can be concluded from the Jarzynski's equality \cite{Jarzynski}, that a crossing point between work distributions of forward and reverse processes is equivalent to free energy difference $\Delta$G between the resulting states and has been used in various studies of mechanical stability and folding/unfolding dynamics of biopolymers \cite{alemany2015free,fox2003using}. The approach allows us to determine $\Delta$G between stretched and relaxed structures of the {\it neck linker} and {\it Ala-only} chains. It is difficult to {\it a priori} ascertain the possible range of work values in such an experiment, since it depends on several factors, such as the experiment's duration, the stretching force's value, the way that force is applied, etc. Nevertheless, we can infer from examples of similar experiments, like the one performed on DNA hairpins \cite{Ritort}, that one should expect work values of order of hundreds of $k_BT$ (where $k_BT$ can be given as approximately 4.14 $pN*nm$). As we can see in  FIG.\ref{work_distr} - work distributions derived from ensembles of stretched and relaxed {\it neck linker} structures are well separated and exhibit larger average values than those typical for the {\it Ala-only} chains. The free energy difference between two equilibrated states $\Delta G$ can be identified as  $\Delta G=1462.4 k_BT$ for the
{\it neck linker} and $\Delta G= 882k_BT$ for the {\it Ala-only} chain. These rough estimates of $\Delta G$, used in conjunction with the Aarhenius definition of the rate constant, seem to imply that the {\it neck linker} not only stretches more effectively than the arbitrarily chosen polyalanine peptide but also returns faster to the relaxed conformation. This conclusion is in line with its documented greater elasticity.

\begin{figure*}[!htb]
\begin{tabular}{c}
\includegraphics[angle=0,width=1\columnwidth]{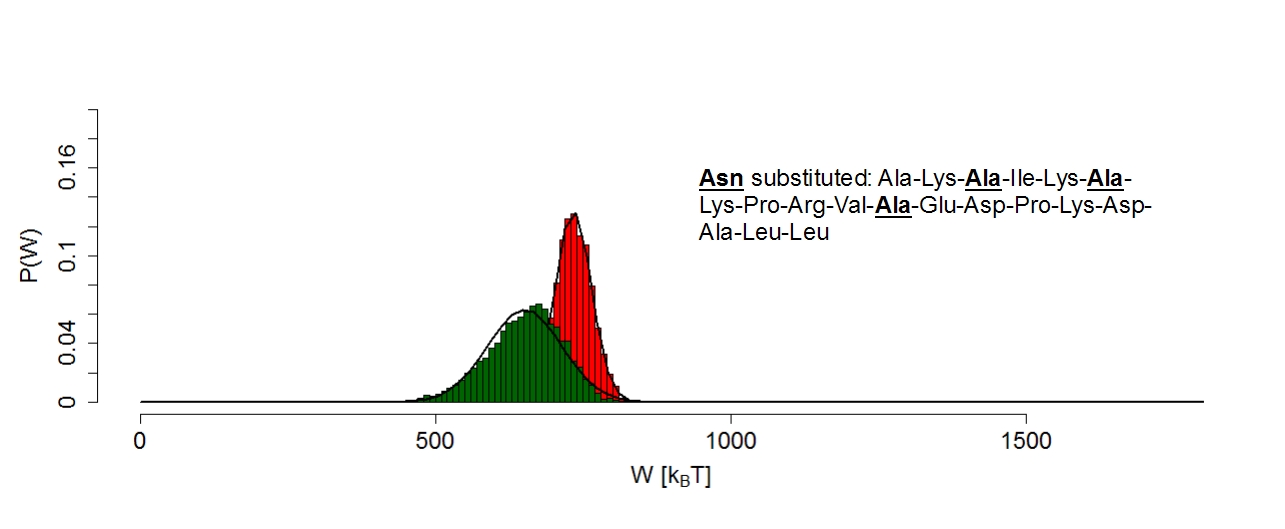} \\
\includegraphics[angle=0,width=1\columnwidth]{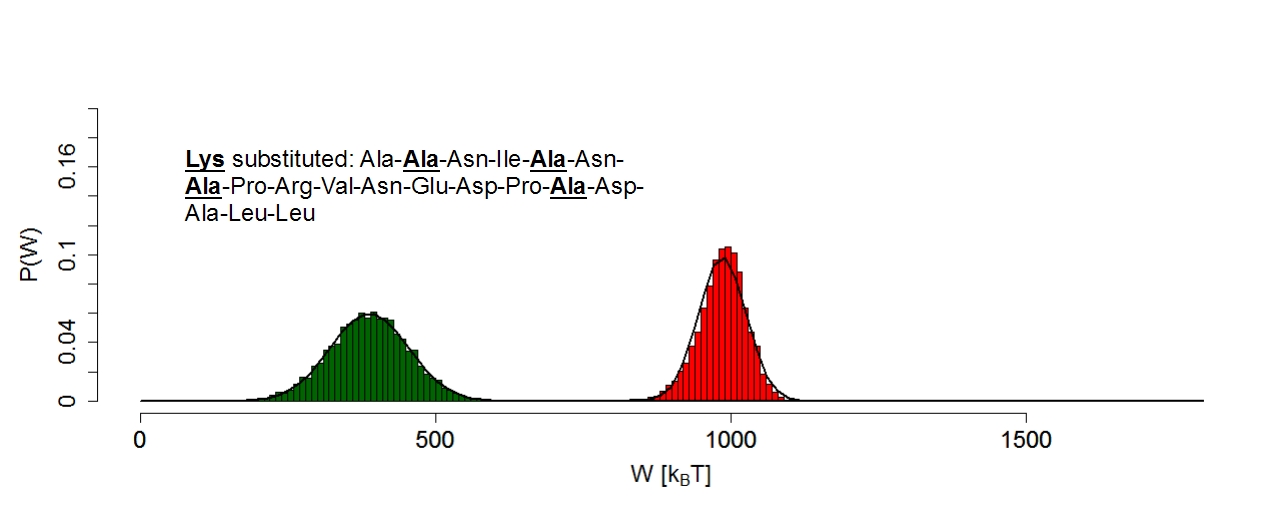} \\
\includegraphics[angle=0,width=1\columnwidth]{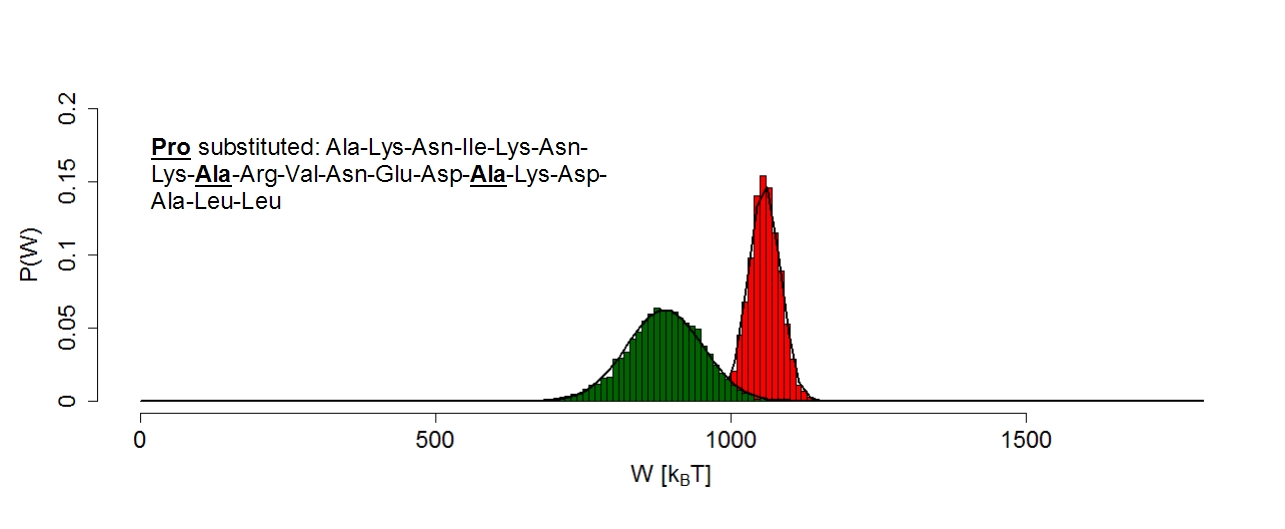} \\
\end{tabular}
\caption{
Reconstruction of probability density functions (PDFs) of work done on the {\it end-to-end} distance in pulling {\it in-silico} experiments on the modified linker chains (the stretching force value set at 1300 $kJ*mol^{-1}*nm^{-1}$). The Gaussian curves are fitted to the derived histograms of the {\it no-Asparagine} peptide (top), the {\it no-Lysine} peptide (middle) and the {\it no-Proline} one (bottom). The ensembles of stretched structures are red colored, while the ensembles of relaxed structures are green.
}
\label{mutant_w_distr}
\end{figure*}

Distributions displayed in FIG.\ref{mutant_w_distr} suggest that the elasticity of all 3 modified {\it neck linker} sequences suffered, compared to the original one (see FIG.\ref{work_distr}). Just like in  case of the polyalanine peptide, the value of work performed on them hardly crosses the point of $1000 k_BT$, while that of the {\it neck linker} easily passes the $1500 k_BT$ mark. Additionally, the removal of Asparagine from the chain has influenced its ability to spontaneously retract most severely, both that ability and its stretching are less effective than that of the polyalanine peptide. The substitution of Proline seems to result in similar behavior to the {\it Ala-only} chain. The {\it no-Lysine} chain's ability to retract seems to even surpass that of the original chain. All this may hint at Asparagine being a crucial part, when it comes to retracting during relaxation process. Asparagine side chain consists of sole amine group, which may have a stabilizing effect through its ability to partake in hydrogen bonding. Lysine side chain contains amine group as well, it is however preceded by a conventional chain of 4 methylene groups, a fact that probably keeps the amine group away from the peptide's backbone. Proline may be adding to the stabilising effects, as well as providing an extra push to the stretching ability with its conformational changes.

\begin{table*}[ht]
\centering
\begin{tabular}{|c|c|c|c|}
\hline
Peptide type & $<W>_{stretch}$ [$k_BT$] & $<W>_{relax}$ [$k_BT$] & $\Delta$ G [$k_BT$] \\
\hline
{\it neck linker} & $1569.9 \pm 24.9$ & $1211.6 \pm 68.2$ & $1462.4 \pm 0.1$ \\
\hline
{\it Ala-only} & $949.7 \pm	23.9$ & $696.0 \pm 81.5$ & $882 \pm 0.1$ \\
\hline
{\it no-Asparagine} & $733.7 \pm	30.5$ & $651.0 \pm	63.3$ & $690.5 \pm 0.1$ \\
\hline
{\it no-Lysine} & $985.6	\pm 40.4$ & $388.0 \pm 66.7$ & $752.8 \pm 0.1$ \\
\hline
{\it no-Proline} & $1056.8 \pm 26.8$ & $888.0 \pm	63.9$ & $997.6 \pm 0.1$ \\
\hline
\end{tabular}
\caption{
Distribution means and $\Delta$G between stretched and relaxed chains of the original {\it neck linker}, the {\it Ala-only} peptide and for 3 modified (mutant-like) structures, with the stretching force value preset to 1300 $kJ*mol^{-1}*nm^{-1}$.
}
\label{dG_val}
\end{table*}

\begin{figure*}[!htb]
\begin{center}
\resizebox{1\columnwidth}{!}{
\includegraphics{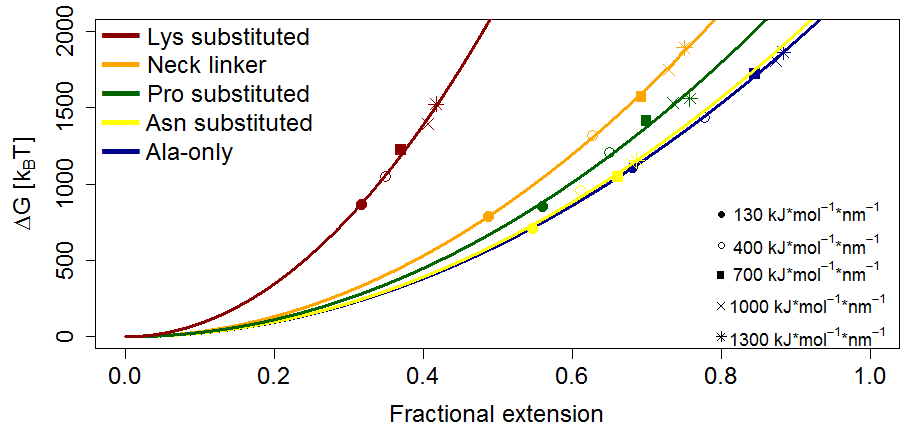}

} 

\caption{
$\Delta$G expressed as a function of the fractional extension for the {\it  neck linker}, {\it Ala-only} peptide, {\it no-Asparagine} chain ({\it Asn substituted}), {\it no-Lysine} chain ({\it Lys substituted}) and {\it no-Proline} chain ({\it Pro substituted}). Several different $\Delta$G values have been acquired by repeating the {\it in silico} experiments while using different values of constant stretching force equal to 130, 400, 700, 1000 and 1300 $kJ*mol^{-1}*nm^{-1}$. Parabolic curves have been fitted to the plots. Quadratic coefficients of fitted plots are: Lys substituted: $8.68*10^3$; Pro substituted: $2.81*10^3$; Asn substituted: $2.45*10^3$; Neck linker: $3.32*10^3$; Ala-only: $2.39*10^3$ and represent the effective stiffness (effective spring constant) of the stretched structure.
}
\label{dG_fraction}
\end{center}
\end{figure*}

The data gathered in remaining in-silico experiments (stretching force values: 130, 400, 700 and 1000 $kJ*mol^{-1}*nm^{-1}$) have been used together with the data for the stretching force value of 1300 $kJ*mol^{-1}*nm^{-1}$) showcased in previous figures. In FIG.\ref{dG_fraction} the relation between fractional extension and $\Delta$G has been shown, with all chains and force values included. 
The quadratic curves fitted to the data points confirm harmonic spring-like behaviour of simulated chains and stay §in line with analysis of single polymer dynamics presented elsewhere (see e.g.Ref.\cite{Latinwo}). The {\it  neck linker} curve is characterised by a quadratic coefficient of a greater value than all other curves, save one. In fact, the {\it  neck linker} curve and the {\it polyalanine} (the {\it Ala-only} peptide) curve seem to define a range of coefficient values that go from the least unique chain (the {\it Ala-only} peptide) to the one present in properly functioning biostructures (the {\it  neck linker}). Predictably, the least unique chain is also the least effective spring, while the other one is most effective. The curves depicting the characteristics of {\it no-Asparagine} chain and {\it no-Proline} chain fall in between these two extremes. It could thus suggest that chains decrease in effectiveness towards the {\it polyalanine} with the removal of {\it Proline} and {\it Asparagine} residues. The lack of {\it Asparagine} in sequence seems to impact the elasticity more profoundly, with {\it no-Asparagine} curve running very close to the {\it polyalanine} curve. This result agrees with our previous conclusions. Intriguingly, the {\it no-Lysine} curve is to the far left of the other curves, including {\it  neck linker} curve. Its quadratic coefficient value reflects this difference: it is close to $8.7$ while the coefficients of 4 other curves have got values ranging from 2.3 to 3.4 ($10^3 k_BT$). Judging from this, we may argue that the substitution of {\it Lysine} residues actually improves the spring-like performance of the chain.
It is possible to draw the conclusion from this particular result that the substitution of {\it Lysine} residues may improve neck linker's performance within the context of kinesin motor's movement along microtubules. However, it is also possible to assume that there is an optimal range within which biologically viable springs operate and that such drastic surge of elasticity takes {\it no-Lysine} chain outside of this range. If it were to be so, then it is probable that this optimal biological range coincides with the range defined by {\it  neck linker} and {\it polyalanine} chains, as shown in the FIG.\ref{dG_fraction}. Indeed, further inquiries may prove enlightening as to whether {\it  neck linker} chain corresponds to the optimal structure in the context of, first, kinesin protein, then whole group of molecular motors and, finally, in the context of all protein native structures.


\section*{Conclusions}

Pulling experiments on single-molecules provide a quantitative characterization of unfolding and relaxation mechanisms of biomolecules. Despite many nanotechniques like atomic force microscopy, laser tweezer or fluorescence resonance energy transfer are available today and used in combined protocols, they may not reveal molecular mechanisms underlying modulation of protein's elasticity, especially under costly conditions of manipulating local mutations of investigated molecules. In order to overcome these difficulties, mechanical models of molecular dynamics can be used as guiding insight into consequences of local modifications of protein structure on its elasticity and response to external mechanical stress \cite{Mei,Chong,Kutys,Harris}. Computational all-atoms MD or coarse-grained MD simulations on single molecule pulling experiments are frequently a complementary tool in analysis of entropic elasticity of polymers or protein molecules and facilitate development and design of single-molecule force spectroscopy \cite{Block,Latinwo,Ritort,hariharan2009insights}. 

Notably, entropic forces have been rarely discussed in the context of nanomechanical devices, although entropy-functional units might be easier controllable by external parameters (like temperature or external fields) and their motion induced without changing the chemical structure of the components. Therefore investigations and a design of such entropy-driven systems seems to be of particular interest in the field of engineering of artificial motors for nanoscale transport.

Here we have investigated force response in the structure of kinesin focusing on elastic properties of a spring connecting two separate domains (heads) of the motor protein.
Due to small molecular size of the motor, in most biological applications the viscous forces are overtaking the inertia effects and the overdamped dynamics is a valid approximation \cite{Howard,Linke,BierPRL}. In contrast, our numerical simulations follow the procedure in which full set of Newton's equations of motion are solved to propagate in time the coordinates of all atoms of the structure under considerations.

Presented study {adds} to the notion that the {\it neck linker} regions possess mechanical properties not found in an arbitrary amino acid chain. The {\it neck linker's} models modified with point mutations clearly exhibit different responses to stretching force in comparison to the intact, original structure.
At this point, it still remains unclear, how much those differences depend on amino acid type or on the number of residues being substituted and whether the placement within the sequence factors in. It is also impossible to assert, whether the amino acid types' impacts on {\it neck linker} properties are merely additive, or if the certain residues' combinations produce more nuanced interactions.
{Future studies, both experimental and  theoretical, should shed light on effects of perturbing motor proteins by introduced point mutations and on adaptation of the resulting stepping mechanism to those changes.}

\section*{Data Availability}
The file containing initial atom coordinates of the {\it neck linker} region, used in the beginning of the simulations' setup, is available at \url{https://www.rcsb.org/structure/3b6u}. The GROMACS package employed to execute simulations can be obtained from \url{http://www.gromacs.org/Downloads}. As for the datasets generated and analysed during the current study, they are available from the corresponding author on reasonable request.

\bibliography{sample}

\section*{Acknowledgements}

{This project has been supported in part by the grant from National Science Center 2014/13/B/ST2/02014.}

\section*{Author contributions statement}

M.Ś. and E. G-N. conceived and designed the research. SMD simulations were conducted by M.Ś. and followed by statistical analysis and interpretation of the results performed together with E. G-N. Both authors (M.Ś. and E. G-N.) wrote the paper.

\section*{Competing interests}
The authors declare no competing interests.

\section*{Corresponding author}
Correspondence to {Micha{\l} {\'S}wi\c{a}tek}.

\end{document}